# Polarizing an antiferromagnet by optical engineering of the crystal field

A.S. Disa[1,2*], M. Fechner[1], T.F. Nova[1], B. Liu[1], M. Först[1], D. Prabhakaran[3], P.G. Radaelli[3], A. Cavalleri[1,2,3*]

[1]*Max Planck Institute for the Structure and Dynamics of Matter, Hamburg, Germany*
[2]*The Hamburg Centre for Ultrafast Imaging, Hamburg, Germany*
[3]*Clarendon Laboratory, Department of Physics, Oxford University, Oxford, UK*

**Strain engineering is widely used to manipulate the electronic and magnetic properties of complex materials[1,2]. An attractive route to control magnetism with strain is provided by the piezomagnetic effect, whereby the staggered spin structure of an antiferromagnet is decompensated by breaking the crystal field symmetry, which induces a ferrimagnetic polarization. Piezomagnetism is especially attractive because unlike magnetostriction it couples strain and magnetization at linear order[3], and allows for bi-directional control suitable for memory and spintronics applications[4,5]. However, its use in functional devices has so far been hindered by the slow speed and large uniaxial strains required. Here, we show that the essential features of piezomagnetism can be reproduced with optical phonons alone, which can be driven by light to large amplitudes without changing the volume and hence beyond the elastic limits of the material. We exploit nonlinear, three-phonon mixing to induce the desired crystal field distortions in the antiferromagnet $CoF_2$. Through this effect, we generate a ferrimagnetic moment of 0.2 $\mu_B$ per unit cell, nearly three orders of magnitude larger than achieved with mechanical strain[6].**

*corresponding authors

The control of magnetism via strain is most commonly based on magnetostriction. This effect relates an induced magnetization to the square of the applied stress and is restricted to systems with an equilibrium magnetization, like ferromagnets. Because the coupling is quadratic, the magnetic moment is only modulated in one direction when no external field is present.

Piezomagnetism, on the other hand, is a property of certain antiferromagnetic materials, which manifests as a linear coupling between an applied stress $\sigma$ and an induced magnetization $M$ of the form $M_i = \Lambda_{ijk}\sigma_{jk}$, where $\Lambda$ is the piezomagnetic tensor [3, 7]. Hence, by exploiting the piezomagnetic effect, one can induce a magnetization of either sign from a crystal that possesses no net magnetization (see Fig. 1A).

$CoF_2$ is one of the simplest known piezomagnetic crystals. It has a rutile crystal structure with a body-centered cubic arrangement of cobalt ions, each surrounded by a fluorine octahedron with a tetragonal crystal field [8, 9]. Below $T_N$ = 39K, it is a fully compensated type I antiferromagnet with easy-axis anisotropy, such that the magnetic moments on the Co ions at the center and corner of the cube point along the tetragonal $c$ axis with opposite sign (magnetic space group $P4_2'/mnm'$, see Fig. 1A) [10, 11]. Piezomagnetism in $CoF_2$ arises from the site-selective modification of the Co crystal fields [6, 12]. Strain along the [110] direction distorts the crystal structure in such a way that the in-plane Co-F bond is shortened for the central sublattice and lengthened for the corner one (or vice-versa). As result, the relative energy splitting $\Delta$ of the Co $t_{2g}$ orbitals is decreased on one site and increased on the other (Fig. 1B). In the $d^7$ configuration, the orbital magnetic moment scales as $m_L \sim \Delta^{-1}$; hence, the strain-induced distortion uncompensates the antiferromagnetically ordered moments and generates a net ferrimagnetic polarization. By changing the direction of the strain from tensile to compressive, the direction of the crystal field distortion reverses, and with that the sign of the induced magnetization.

The essential microscopic ingredient in the piezomagnetic effect is the antiparallel distortion of the Co-F bonds on each sublattice. Hence, it could also be induced also through the displacements of certain optical phonons alone, with the advantage of preserving the cell volume. New optical devices enabling resonant driving of lattice vibrations at mid-infrared and terahertz frequencies open up the possibility to selectively control the structure of solids with light [13, 14, 15, 16]. At large phonon amplitudes, exploit lattice nonlinearities can be exploited to rectify atomic displacements and coherently mimic the effect of heteroepitaxial or externally applied strain [17, 18]. In particular, it was recently suggested that a rectification of the $B_{2g}$ Raman phonon in $CoF_2$ would produce the same relative bond displacements as [110] uniaxial strain [19], at constant unit cell volume.

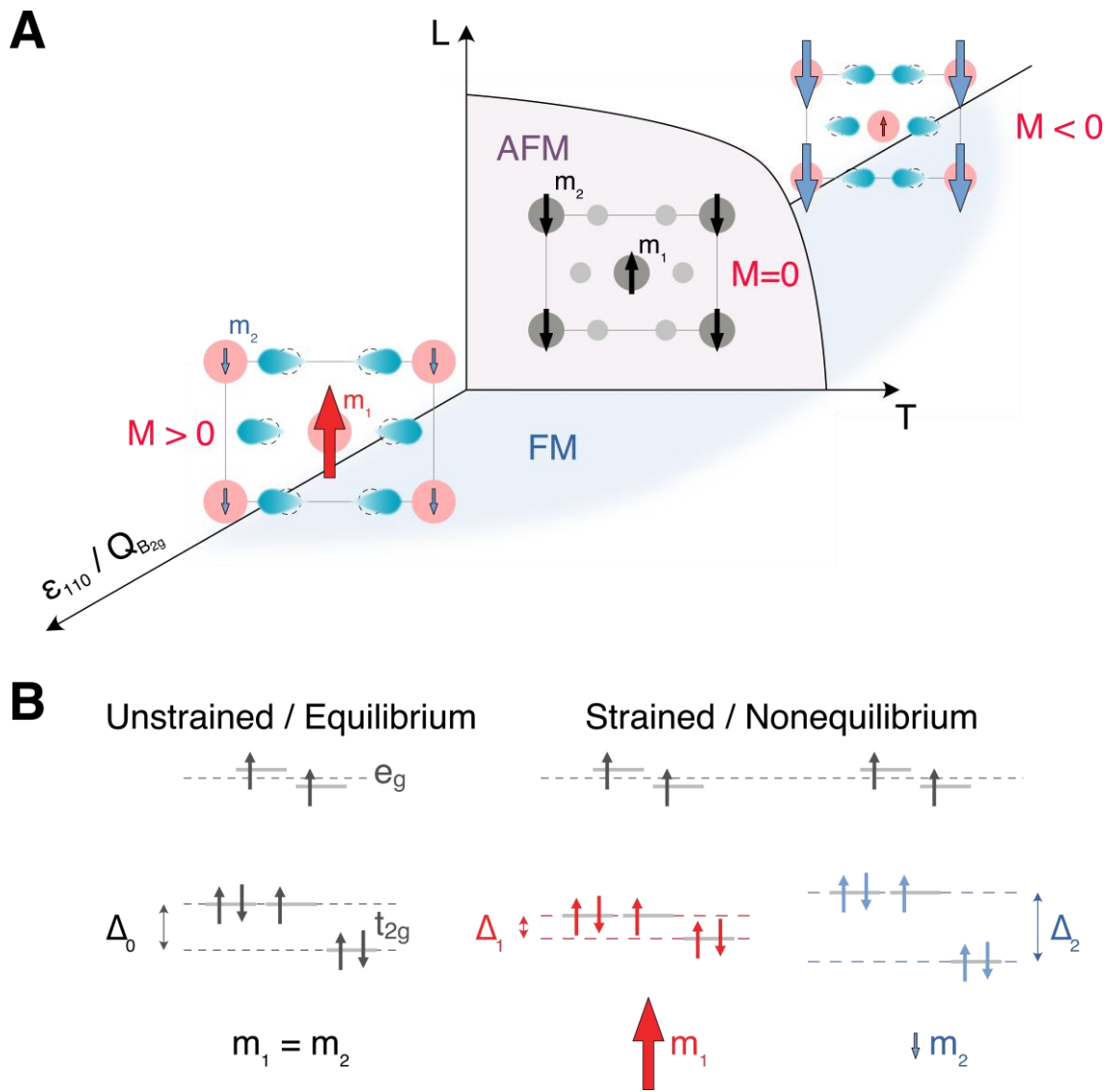

**Figure 1. Piezomagnetic effect in CoF₂.** (A) Under no strain, $CoF_2$ is antiferromagnetic below $T_N$. Under appropriate lattice deformation the antiferromagnetically aligned moments on the Co sites ($m_1$ and $m_2$) uncompensate, leading to ferrimagnetic order and non-zero magnetization ($M$). The lattice distortion (depicted in the $ac$ plane) can be created by uniaxial strain along [110] ($\varepsilon_{110}$) or by atomic displacements along the $B_{2g}$ Raman phonon ($Q_{B_{2g}}$). (B) The $d$ electron crystal field levels for $CoF_2$ showing the origin of the piezomagnetic moment. The energy splitting between the $t_{2g}$ levels ($\Delta$) determines the orbital contribution to the magnetic moment ($m_L \sim \Delta^{-1}$). In the undistorted structure, the energy splitting between the $t_{2g}$ levels ($\Delta$) is the same for the two Co sites (left). When distorted, $\Delta$ shrinks for one site and grows for the other, hence changing the relative magnitudes of $m_1$ and $m_2$.

Here, we validate this approach and generate light-induced ferrimagnetic order in CoF$_2$ by using anharmonic coupling of three volume-preserving optical phonons to drive symmetry-breaking lattice distortions. In contrast to the case of applied strain, this *dynamical piezomagnetic effect* creates orders-of-magnitude larger displacements on time scales not limited by the speed of sound.

The desired lattice distortion corresponds to atomic motions along the coordinates of a Raman-active phonon of $B_{2g}$ symmetry. Hence, it cannot be directly excited by light. Moreover, a purely impulsive excitation, typically achieved using near-infrared pulses, would produce an oscillating vibration about the equilibrium atomic positions and no net displacement of the lattice. Instead, we use the nonlinear coupling between resonantly driven infrared-active (IR) phonons and the Raman (R) mode, which can provide directional atomic motions away from equilibrium.

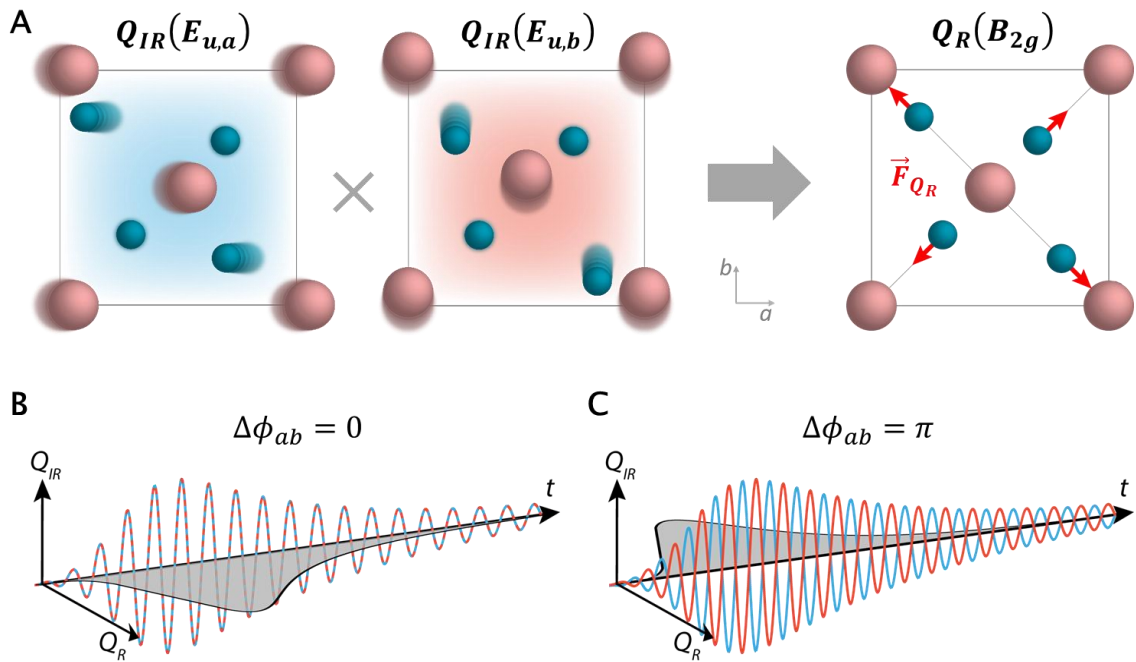

**Figure 2. Breaking symmetry with phonons.** (A) Two orthogonal, degenerate $E_u$ phonons mix nonlinearly to displace the lattice along the $B_{2g}$ Raman mode. The force on the lattice arising from the nonlinear interaction ($\vec{F}_{Q_R}$) is shown in red arrows. (B) The dynamics of the two IR modes (in red and blue) and the Raman mode (in grey) resulting from resonant excitation of both $E_u$ modes by a terahertz pulse. The IR modes oscillate about their equilibrium positions, while the Raman mode is displaced away from equilibrium, creating a time-averaged structural distortion (solid shading). The two IR modes in this case are pumped in phase ($\Delta\phi_{ab} = 0$). (C) The phonon dynamics for the case when the two IR modes are driven out of phase ($\Delta\phi_{ab} = \pi$) showing that the Raman mode is displaced in the opposite direction.

Figure 2 describes the anharmonic lattice interaction used to drive the dynamical piezomagnetic effect [19]. The lowest order phonon coupling is given by the energy term $U = Q_{IR,1}Q_{IR,2}Q_R$, where $Q_{IR}$ and $Q_R$ denote IR and Raman mode coordinates. In the case that $Q_R$ has $B_{2g}$ symmetry, this coupling is symmetry allowed if $Q_{IR,1}$ and $Q_{IR,2}$ both have $E_u$ symmetry. If the two modes are orthogonal and degenerate, this coupling provides a force onto $Q_R$ that follows the envelope of the two optically driven IR modes, leading to a rectified displacement of the lattice along the $B_{2g}$ mode direction (Fig. 2A). The time-averaged displacement of $Q_R$ is proportional to the product of the two IR mode amplitudes, $\langle Q_R \rangle \propto |Q_{IR,a}||Q_{IR,b}|\cos\Delta\phi_{ab}$, where $\Delta\phi_{ab}$ is their relative phase of the two modes. Hence, the lattice displacement is maximal when the two IR modes are driven in phase, and the direction of the displacement flips when one of the two IR modes changes sign (*i.e.* $\Delta\phi_{ab} = \pi$), as shown in Fig. 2B,C. In $CoF_2$, the induced ferrimagnetic moment is linearly proportional to this $Q_R(B_{2g})$ displacement (see Supplement, Sec. S3).

Experimentally, in order to create the required lattice distortion and generate a net magnetization, one must simultaneously drive two degenerate, orthogonal IR modes with $E_u$ symmetry. Figure 3A shows the optical conductivity along the *a* and *b* axes measured on the 100 μm thick $CoF_2$ crystal used in this experiment. The three conductivity peaks identify different $E_u$ phonons with resonant frequencies close to 6, 8 and 12 THz. We resonantly and simultaneously drove the two IR modes at 12 THz (atomic motions also shown in Fig. 3A), for which the largest anharmonic coupling to the $B_{2g}$ Raman mode was predicted (see Supplement, Sec. S3).

Intense THz excitation pulses were generated in a specially designed optical device that is based on two optical parametric amplifiers, a chirping stage, and a difference frequency generation stage using an organic crystal[20], as depicted in Fig. 3B. We produced 500 fs long pulses with peak electric fields up to ~10 MV/cm and a center frequency of ~12 THz (see Supplement, Sec. S1). To excite both *a* and *b* axis modes simultaneously, the linearly polarized pump pulses were oriented at 45° with respect to the crystallographic axes (*i.e.* oriented along the [110] direction), impinging normally to the (001) face of the crystal. The sign of the induced $Q_R$ displacement could then be flipped by rotating the pump polarization by 90° (along the $[1\bar{1}0]$). The induced magnetization was detected via complementary Faraday rotation and circular dichroism measurements at near-infrared/visible wavelengths.

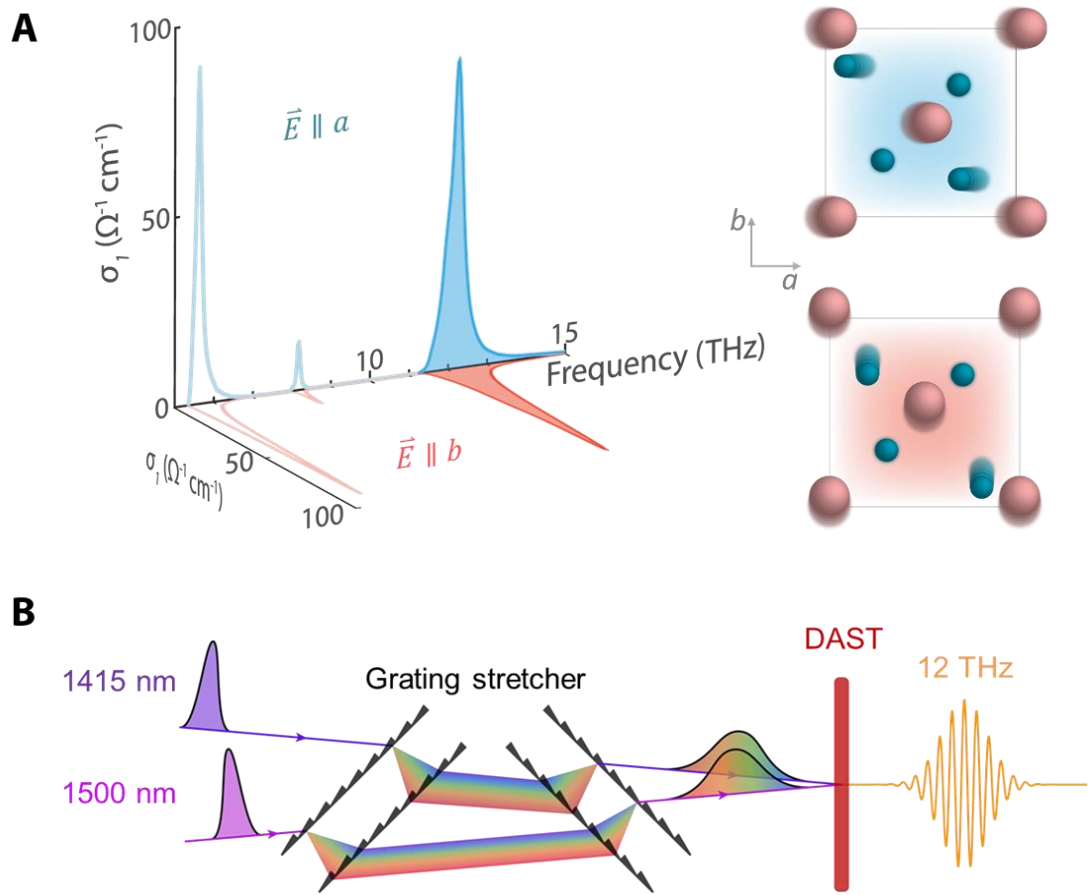

**Figure 3. Driving degenerate infrared phonons in CoF$_2$.** (A) Optical conductivity of CoF$_2$ at 6 K along the two orthogonal in-plane crystal axes (*a* and *b* in blue and red, respectively). The three peaks arise from the doubly degenerate ($E_u$) IR phonons. We pump the mode at 12 THz (solid shading). A top view of the atomic motions of the two components of the pumped mode is shown on the right. (B) Depiction of chirped pulse difference frequency generation for the pump pulses used in this experiment. Two near-IR pulses generated from co-seeded optical parametric amplifiers are stretched using two transmission grating pairs and combined non-collinearly on the organic crystal DAST (4-N,N-dimethylamino-4'-N'-methyl-stilbazolium tosylate) to produce narrow bandwidth, high-intensity pulses with a center frequency of 12 THz.

Time-resolved Faraday rotation measurements taken at $T$ = 4 K are shown in Fig. 4A. Without the pump, no detectable rotation was observed. After photo-excitation, a pump-induced magnetization signal developed, which switched sign after ~7 ps and continued to grow in magnitude before reaching a maximum after ~200 ps. Rotating the pump polarization by 90° reversed the sign of the Faraday rotation, as expected by the anharmonic coupling mechanism discussed above. As a comparison, we conducted the same experiment on a ZnF$_2$ crystal, which has a nearly identical phonon spectrum to CoF$_2$, but is non-magnetic [21, 22]. The ZnF$_2$ crystal showed only a prompt polarization rotation (arising from pump-induced birefringence) and no long-lived signal, suggesting that the signal observed in CoF$_2$ is of magnetic origin.

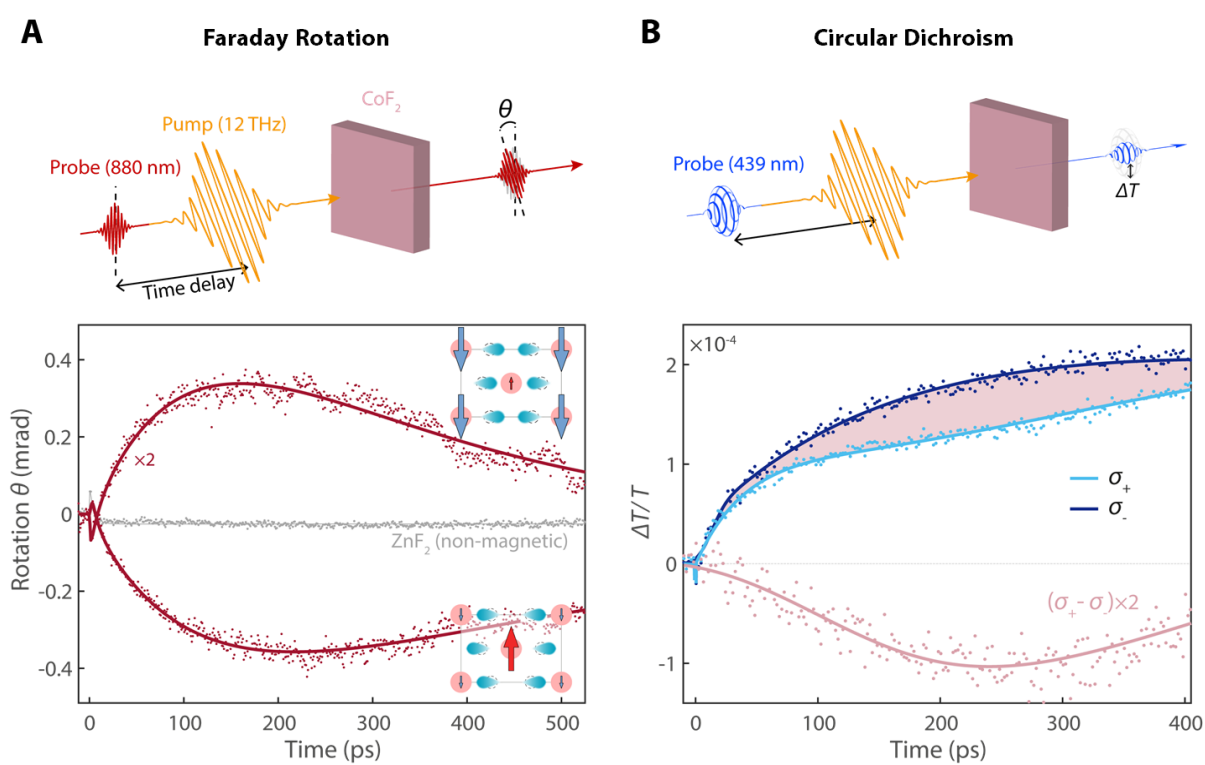

**Figure 4. Time-resolved magneto-optical measurements.** (A, top) Depiction of the THz pump-Faraday rotation probe measurement setup. The 12 THz pump was oriented at 45° with respect to the crystallographic *a* and *b* axes and the polarization rotation angle of a subsequent 880nm probe pulse was detected. (bottom) Faraday rotation data on CoF$_2$ (red) for two polarizations of the pump pulse (+45° and -45°) as a function of time delay. They display similar time dependences, with a small signal at short times that reverses, saturates and eventually decays back to zero. The inset shows the induced structural distortion (in the *ac* plane) and resulting ferrimagnetic state for the two pump polarizations. The same measurement taken on ZnF$_2$ is plotted in gray, exhibiting no long-lived response. (B, top) Depiction of the THz pump-circular dichroism probe measurement setup. The absorption of left and right circularly polarized probe pulses at 439 nm is detected as a function of time delay. (bottom) The relative change in transmission for left (dark blue) and right (light blue) circular polarized probe pulses. The shaded area depicts the circular dichroism, which is plotted on a 2× scale in pink. The time dependence of the circular dichroism matches that of the Faraday rotation. In all panels, dots are measured data points and solid lines are double-exponential fits.

In order to confirm the magnetic nature of the signal, we also carried out time-resolved circular dichroism measurements (Fig. 4B). At equilibrium, circular dichroism has been observed in $CoF_2$ near magnetically active electronic transitions in the presence of a large externally applied magnetic field, with the strongest effect found at the 439 nm absorption line[23, 24]. We probed the pump-induced change in transmission ($\Delta T/T$) at this wavelength using left and right circularly polarized pulses. A large difference in the relative $\Delta T/T$ was detected for the two polarizations after the THz pump excitation, with a time scale matching that of the Faraday rotation measurements.

Taken together, the observations reported above evidence a non-equilibrium state with a net magnetization and a response time of approximately 100 picoseconds.

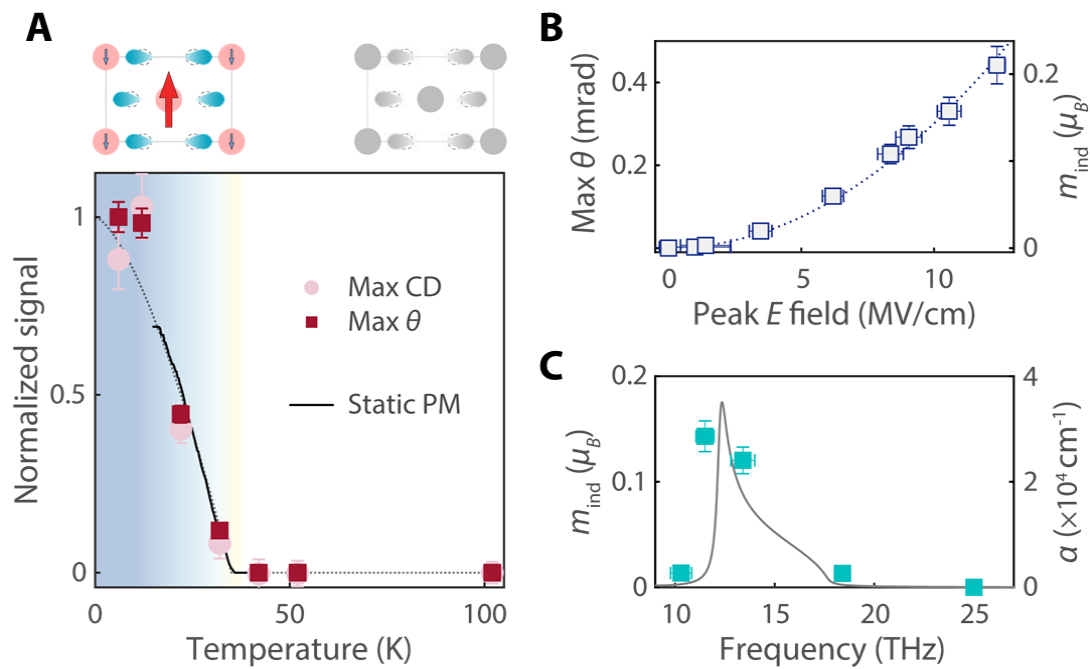

**Figure 5. Characterization of the pump-induced state.** (A) Temperature dependence of the maximum magnitude of the pump-induced Faraday rotation angle and circular dichroism (at ~200 ps), normalized to its value at 4 K. The static piezomagnetic response is shown in solid black (adapted from Ref. 25), and the dotted line is a guide to the eye. The insets at the top show the dynamical piezomagnetic state in the $ac$ plane below $T_N$, and the distorted (but not magnetic) state above $T_N$. (B) Pump peak electric field dependence of the maximum Faraday rotation angle and the associated induced magnetic moment per unit cell at the sample surface. (C) Pump frequency dependence of induced moment per unit cell for a peak field of 10 MV/cm (solid squares) and extinction coefficient of the phonon (solid gray).

Figure 5A shows the magnitude of the pump-induced Faraday rotation and circular dichroism as a function of temperature. Both signals reduced upon warming the sample from 4 K and vanished above 40 K, coinciding with the Néel temperature $T_N$. The temperature evolution of the non-equilibrium magnetization followed that of the static piezomagnetic response of CoF$_2$ (Ref. 25). In both cases, the magnitude of the induced ferrimagnetic polarization should scale with the staggered magnetization of the equilibrium antiferromagnetic state.

The dependence of the Faraday rotation signal on the peak electric field of the pump pulse is shown in Fig. 5B. Using the Verdet constant for CoF$_2$ [26], we determined the induced magnetic moment at the sample surface $m_{ind}$ and found that it scales as the square of the pump electric field (see Supplement, Sec. S2). Since the electric field linearly excites the two IR modes ($Q_{IR}^a$ and $Q_{IR}^b$) and we expect $m_{ind} \propto Q_R$, this quadratic field dependence is consistent with the three-phonon coupling ($Q_{IR}^a Q_{IR}^b Q_R$) described in Fig. 3. Moreover, as a function of pump wavelength, the magnitude of the non-equilibrium moment showed a resonant enhancement at the eigenfrequency of the $E_u$ phonon, as shown in Fig. 5C, demonstrating that the magnetization dynamics are indeed driven by the excited optical phonons.

We also found that at the maximum pump electric field of 12 MV/cm, the induced magnetic moment reached $m_{\text{ind}} \approx 0.2 \, \mu_B$ per unit cell, in agreement with predictions from first-principles calculations (Fig. S4). This value is more than 400 times greater than the largest reported piezomagnetic moment induced statically in CoF$_2$ [6]. From the known piezomagnetic coefficients, a uniaxial stress of 40 GPa would be required to generate an equivalent moment, whereas typical tensile strengths of fluoride and oxide single crystals are < 0.5 GPa [27]. Since our pump excitation selectively distorted the $Q_{B_{2g}}$ mode without deforming the unit cell, much larger atomic displacements could be imposed than are possible with static strain (up to 2% of the equilibrium bond length; see Supplement, Sec. S3).

We now turn to a discussion of the dynamic behavior of the light-induced magnetization. A minimal model can be obtained from a free-energy description including the coupling between the antiferromagnetic order parameter (*L*), the induced magnetization (*M*), and the *B$_{2g}$* structural distortion ($Q_R$) (detailed in Supplement, Sec. S4). In this picture, the system initially occupies a minimum in the energy landscape where $M = 0$ and $L \neq 0$ (Fig. 6A). The magnetization dynamics are activated by the $B_{2g}$ phonon displacement, which shifts the energy minimum to finite $M$ and back on the time scale of the phonon lifetime (~5 ps). The subsequent evolution of the magnetic system after this prompt stimulus can be described as an overdamped oscillator: the kinetic energy imparted by the drive leads to the inertial growth of $M$, and at longer times the generated angular momentum equilibrates through spin-lattice relaxation processes [28, 29]. Figure 6B shows the computed $M(t)$ from this model for different values of the driving electric field, which reproduce well the experimental features.

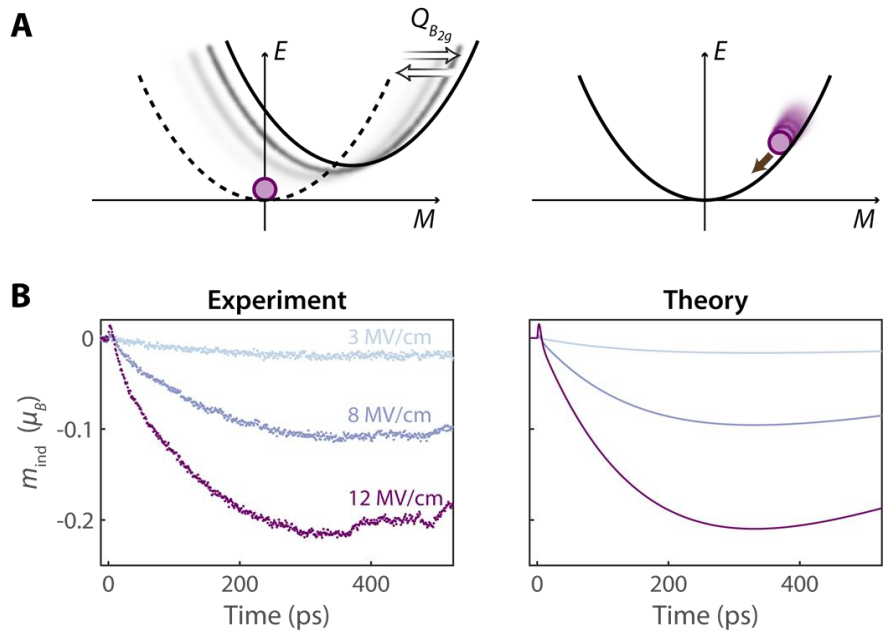

**Figure 6. Pump-induced magnetization dynamics.** (A,left) Illustration of the shift of the free-energy minimum to non-zero magnetization ($M$) upon displacing $Q_{B_{2g}}$, and (right) the subsequent evolution of the system in the equilibrium energy landscape. (B) Comparison between experiment and theory for the time dependence of the induced magnetic moment per unit cell for different pump electric field strengths. The theoretical calculations are obtained from the model described in Supplementary Sec. S4 and depicted in (A).

The proposed model provides an intuitive description of the experimentally observed magnetization dynamics. Microscopically, the dynamics of the magnetic moments of each sublattice are described by coupled Landau-Lifshitz-Bloch equations [30, 31]. In this framework, the observed growth and decay may arise from the slow longitudinal relaxation of $M$ after the impulsive drive, with a time scale associated with the equilibration of the moments on the two sublattices. An outstanding question is the transfer mechanism between the prompt *g*-factor imbalance (created by the phonon) and the moment imbalance observed at later times (see Supplement, Sec. S4). One possibility is that the $B_{2g}$ mode displacement modulates the anisotropy and exchange fields, which would split the magnon bands and could lead to the observed time-dependent behavior. Another possibility is that the transient lattice distortion is reinforced by the induced magnetization, creating a metastable state with slow dynamics, as has been observed in light-induced ferroelectrics [18].

The demonstrated phonon-driven, ultrafast analog of the piezomagnetic effect provides a new mechanism to manipulate magnetism in antiferromagnetic systems [32]. This dynamical effect circumvents typical limits of strain control, offering possibilities to explore novel, out-of-equilibrium phase behavior of correlated materials (like unconventional superconductors[1,2]) and generate light-induced functional responses under extreme lattice deformations. Importantly, this approach differs from previous optical [33] and phononic [15,16] methods of magnetic phase control, as it represents the rare case of a phase transition driven by coherent light-induced breaking of lattice symmetry.

## Acknowledgments

We thank J. Chen for help preparing the samples and assistance with the optical experiment. This work received funding from the European Research Council under the European Union's Seventh Framework Programme (FP7/2007-2013)/ERC [grant agreement no. 319286 (QMAC)] and the Cluster of Excellence 'CUI: Advanced Imaging of Matter' of the Deutsche Forschungsgemeinschaft (DFG) - EXC 2056 - project ID 390715994. Work done at the University of Oxford was funded by EPSRC Grant No. EP/M020517/1, entitled Oxford Quantum Materials Platform Grant. A.S.D. was supported by a fellowship from the Alexander von Humboldt Foundation.

# Supplementary materials for "Polarizing an antiferromagnet by optical engineering of the crystal field"

A.S. Disa[1,2], M. Fechner[1], T.F. Nova[1], B. Liu[1], M. Först[1], D. Prabhakaran[3], P.G. Radaelli[3], A. Cavalleri[1,2,3]

[1]Max Planck Institute for the Structure and Dynamics of Matter, Hamburg, Germany
[2]The Hamburg Centre for Ultrafast Imaging, Hamburg, Germany
[3]Clarendon Laboratory, Department of Physics, Oxford University, Oxford, UK

## S1. Experimental set-up

The THz pump-optical probe set-up for this experiment is shown in Fig. S1a. We used the output of a Ti:sapphire amplifier (800 nm wavelength, 100 fs duration, 5 mJ pulse energy, 1 kHz repetition rate) to pump two optical parametric amplifiers (OPA). The signal output of each OPA had roughly 350 µJ pulse energy and 75 fs pulse duration with independently tunable center wavelengths in the near-infrared.

The THz pump pulses were generated through a chirped difference frequency generation (DFG) process, similar to Ref. [1], by mixing the two OPA signal outputs in the organic crystal DAST (4-N,N-dimethylamino-4'-N'-methyl-stilbazolium tosylate). Before impinging on the DAST crystal, the near-infrared signal pulses were stretched by a transmission grating pair in order to minimize optical rectification and produce narrow bandwidth THz pulses. For the generation of pulses with 12 THz center frequency by DFG, the two signal wavelengths were chosen to be 1500 nm and 1415 nm. For the pump frequency dependence, one signal wavelength was fixed at 1500 nm, while the other was varied to achieve the desired difference frequency. The frequency of the resulting THz pulses was measured by a fourier transform infrared (FTIR) interferometer and their electric field profile was detected via electro-optic sampling in GaSe (Fig. S2). The pump pulses for this experiment had durations of 400-600 fs (~1 THz bandwidth) and maximum pulse energies of ~2 mJ.

The probe pulses for the Faraday rotation measurements were derived from a non-collinear OPA (NOPA) pumped by a portion of the 800 nm laser amplifier output. The NOPA provided < 45 fs long pulses centered around 880 nm wavelength. After passing through the sample, the transmitted probe beam was split by a polarizing Wollaston prism into two orthogonal linear components (P and S). The two polarization components were measured by a balanced Si photodetector to determine the rotation angle.

For the magnetic circular dichroism measurements, the NOPA output was sent through a 2 mm thick BBO (Beta Barium Borate) crystal, phase matched to generate pulses with 439 nm center wavelength and 2 nm bandwidth. Before impinging on the sample, the 439 nm pulses were sent through a linear polarizer and a quarter-wave plate to create circularly polarized pulses (either σ+ and σ-). A photomultiplier tube was used to detect the intensity of the transmitted beam after the sample, and the relative transmission of the two orthogonal components were compared to extract the circular dichroism.

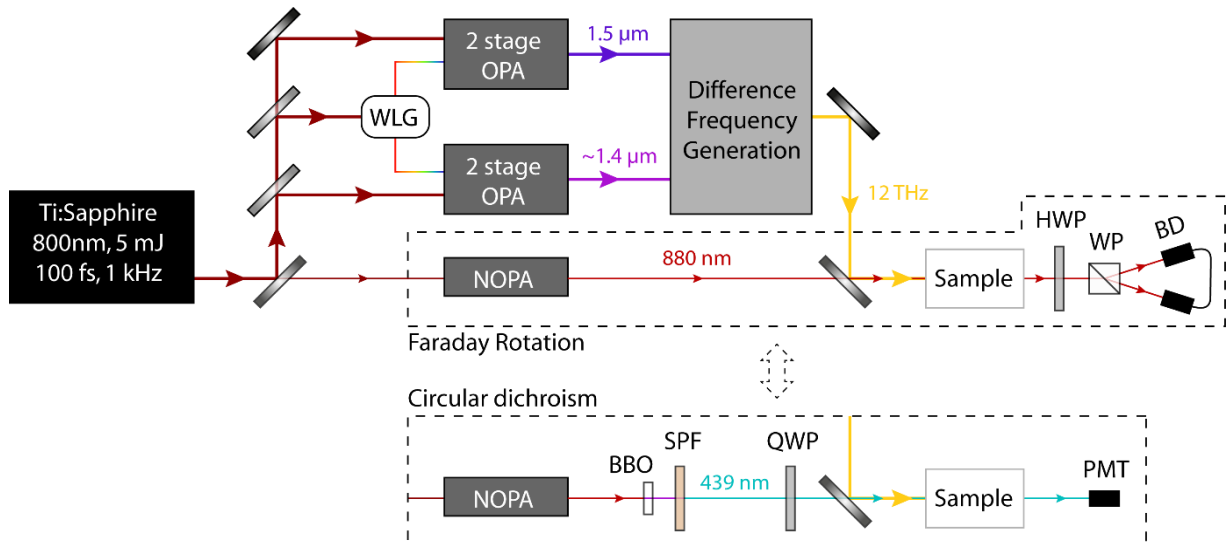

**Figure S1.** Schematic of the optical set-up for the time-resolved measurements of THz pump-induced magnetization. The top panel shows the full generation and detection scheme for the Faraday rotation probe measurements. For the circular dichroism probe measurements, the components in the dashed box in the top panel are replaced by those in the bottom panel. The labeling of the components is as follows: WLG = white-light generation, OPA = optical parametric amplifier, NOPA = non-collinear OPA, HWP = half-wave plate, WP = Wollaston prism, BD = balanced detector, BBO = 2mm thick Beta Barium Borate crystal (Beta Barium Borate), SPF = short-pass filter, QWP = quarter-wave plate, and PMT = photomultiplier tube.

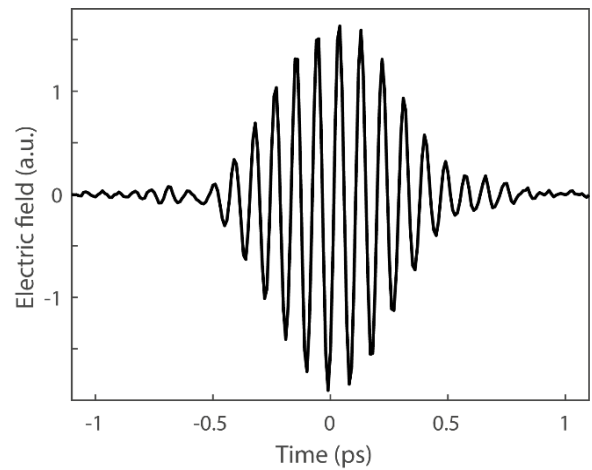

**Figure S2.** Electric field profile of the THz pump pulse measured by electro-optic sampling in GaSe at the sample position. The center frequency is 11.5 THz with a 1 THz bandwidth FWHM. The pulse duration is 500 fs FWHM.

The sample used for the experiment was a 100 µm thick (001)-oriented $CoF_2$ single crystal [2]. The pump and probe pulses were focused onto the sample surface at normal incidence. The THz pump was linearly polarized with the polarization oriented at 45° with respect to the *a* and *b* axes of the crystal (along either the [110] or [1$\bar{1}$0] crystallographic directions). For the Faraday rotation measurements, the probe polarization was orthogonal to that of the pump before the sample. At the sample position, the size of the pump spot was ~70 µm FWHM and the probe size was roughly half that of the pump.

The measurements were carried out in a He-flow cryostat with a base temperature of 4 K. In the antiferromagnetic phase in $CoF_2$, there are two possible orientations of the antiferromagnetic order parameter (180° with respect to each other). In order to reduce the domain distribution, we applied a small (< 1 mT) magnetic field along the c axis while cooling below $T_N$ (= 39 K), which was previously shown to stabilize large volume fractions of single 180° antiferromagnetic domains in MnF$_2$ (Ref. [3]).

## S2. Experimental determination of induced magnetic moment

The magnetic moment induced by the THz pump was determined from the measured pump-induced Faraday rotation angle. The Faraday effect relates the rotation angle $\vartheta$ of linearly polarized light passing through a medium of thickness $L$ to a magnetic flux density $B$ via the relation, $\theta = VBL$, where $V$ is the Verdet constant. In the absence of an external magnetic field, this becomes $\theta = \mu_0 VML$, where $M$ is the magnetization of the medium.

In our experiment, we induced a magnetization which varied as a function of depth because of the finite penetration depth $\delta$ of the pump pulse. The probe penetration depth was much larger than the sample thickness[4, 5] (~1 mm at 880 nm), so the measured rotation angle arises from an integration of the induced magnetization over the entire sample: $\theta = \mu_0 V \int_0^L M(x)dx$. The pump intensity decays exponentially within the material (over the length scale $\delta$), and as shown in Fig. 5B in the main text, the induced $M$ scales linearly with the pump intensity. Then, the induced magnetization at the surface of the material $M_{\text{ind}} = \frac{\theta}{\mu_0 V \delta (1 - e^{-L/\delta})}$.

Using the Verdet constant determined from static magnetic field measurements[6] $V$ = 2.4 °/(cm·G), we find $M_{ind}$ = 28 ± 3 G for the maximum rotation angle of 0.44 ± 0.02 mrad induced by the pump at 12 THz ($\delta$ = 3.8 μm). Taking the volume of the CoF$_2$ unit cell (70.099 Å$^3$, as measured by x-ray diffraction[7]), this induced magnetization corresponds to an induced moment of $m_{ind}$ = 0.21 ± 0.03 $\mu_B$ per unit cell. The values of $m_{ind}$ in Figs. 5B and C were computed in the same manner as above for all pump intensities and wavelengths. The value and uncertainty of the maximum rotation angle were extracted from double exponential fits to the time-dependent Faraday rotation data. The penetration depth was determined from the measured optical conductivity (shown in Fig. 3A in the main text).

## S3. Ab-initio modeling

To gain understanding of the microscopic mechanism and estimate the magnitude of the optically induced ferrimagnetic moment, we performed computations in the framework of density functional theory. To this end, we first compare the magnitude of structural changes induced by shear pressure and optical phonon excitation. In the second step, we then examine the magnetic properties of the distorted crystal structure. The numerical framework and setting of our computations are listed in Sec. S3.5 (Computational Settings).

### S3.1. Equilibrium structural properties

Investigating structural deformations upon strain or displaced phonons requires a force-free reference structure, which was obtained here by structurally relaxing the CoF$_2$ unit cell to the lowest-energy state. As a starting point, we took the tetragonal unit cell of the $P4_2/mnm$ space group from experiment with antiferromagnetic spin-order on the Co atoms. After relaxation, the lattice constants become $a$ = 4.57 Å and $c$ = 3.10 Å. The fluorine atoms, which exhibit only internal structural degrees of freedom at their $f$-Wyckoff positions, are positioned at $x$ = 0.303. All values are in agreement with the experimentally determined structure[7], besides the typically underestimated lattice constants in the local density approximation.

Phonon spectra of the optimized structure were then computed by use of the PHONOPY software package [8]. The zone center modes at q = (0,0,0) are the most relevant for the pressure-induced modulations and phonon excitation, and we list our computed frequencies in Tab. S1. We also computed the phonon dispersion over the full Brillouin zone and found only real frequency phonons, confirming the structural stability of the relaxed $CoF_2$ unit cell. Following the approach of Ref. [9], we finally computed the mode effective charges for the infrared active (IR) phonon modes, which are also listed in Tab. S1.

**Table S1.** Computed phonon frequencies for Raman and infrared active phonon modes of $CoF_2$, together with their mode effective charges.

| Symmetry | Frequency (THz) | Symmetry | Frequency (THz) | $Z^*$ $(e/\sqrt{u}\text{Å})$ |
|---|---|---|---|---|
| $E_g$ | 8.6 | $E_u$ | 5.6 | 0.31 |
| $A_{1g}$ | 12.5 | $E_u$ | 8.6 | 0.06 |
| $A_{2g}$ | 7.3 | $E_u$ | 13.7 | 0.58 |
| $B_{1u}$ | 4.9 | $A_{2u}$ | 11.7 | 0.66 |
| $B_{1u}$ | 12.4 | | | |
| $B_{1g}$ | 1.6 | | | |
| $B_{2g}$ | 16.1 | | | |

## S3.2. Application of static strain

Structural modulations due to static shear strain $\varepsilon$ were calculated by imposing a finite shear pressure $\sigma$ within the $a$-$b$ plane of the $CoF_2$ unit-cell

$$\boldsymbol{\varepsilon} = \boldsymbol{S}\,\boldsymbol{\sigma} = \boldsymbol{S}\,\{0,0,0,\sigma_4,0,0\}.$$

Here, $\boldsymbol{S}$ is the compliance tensor ($\boldsymbol{S} = \boldsymbol{C}^{-1}$) and $\boldsymbol{C}$ the bulk elastic moduli, which were computed for the ground-state structure and are presented in Tab. S2.

We investigated shear stresses between 12.5 and 85.5 GPa and structurally relaxed the internal coordinates of the unit cell to obtain a force-free state. Note that the unit-cell lattice constants were kept fixed during this structural relaxation.

The resulting changes of the internal coordinates were mapped to the eigenvectors of the zone center phonons. We found that the major contribution of the pressure distortion corresponds to the displacements along the coordinates of the $B_{2g}$ phonon mode. Figure S2A shows the calculated amplitudes of the $B_{2g}$ mode as a function of the applied pressure, revealing a linear dependence.

**Table S2.** Elastic moduli of $CoF_2$ in kBar.

| $C_{11}$ | 1464 | $C_{33}$ | 2228 | $C_{44}$ | 1050 | $C_{66}$ | 401 |
|---|---|---|---|---|---|---|---|
| $C_{55}$ | 400 | $C_{12}$ | 1266 | $C_{13}$ | 1087 | $C_{23}$ | 1070 |

## S3.3. Nonlinear phonon dynamics

Dynamic distortions of the crystal structure in response to the THz optical excitation were calculated by following the established approach of nonlinear phononics, developed and applied in Refs. [10, 11, 12]. The relevant anharmonic potential for $CoF_2$ reads

$$V = \frac{\omega_{IR}^2}{2}Q_{IR,1}^2 + \frac{\omega_{IR}^2}{2}Q_{IR,2}^2 + \frac{\omega_R^2}{2}Q_R^2 + c_{12R}Q_{IR,1}Q_{IR,2}Q_R + h.o.t., \quad (S1)$$

where $Q_{IR,1}$ and $Q_{IR,2}$ are IR active modes, $Q_R$ is a Raman active mode, $c_{12R}$ is the trilinear coupling constant considered in Ref. 13, and $h.o.t.$ refers to higher order terms. In Tab. S3, we list the computationally obtained coupling constants $c_{12R}$ for the case when $Q_R$ is the $B_{2g}$ symmetry mode. As discussed in the main text, $c_{12R}$ is only non-zero when $Q_{IR,1}$ and $Q_{IR,2}$ are orthogonal $E_u$ symmetry components ($E_{u,a}$ and $E_{u,b}$). We find that the two $E_u(13.7)$ modes exhibit an exceptionally strong coupling to the $B_{2g}(16.1)$ mode.

**Table S3.** Trilinear coupling constants for $c_{12R}Q_{IR,1}Q_{IR,2}Q_R$ with $Q_R = B_{2g}$(16.1THz). The two components of the doubly degenerate $E_u$ modes are labeled $a$ and $b$ with their frequencies in brackets. The coupling constants are given in units of meV/($\sqrt{u}$Å)$^3$.

|  | $E_{u,a}(5.6)$ | $E_{u,b}(5.6)$ | $E_{u,a}(8.6)$ | $E_{u,b}(8.6)$ | $E_{u,a}(13.7)$ | $E_{u,b}(13.7)$ |
|---|---|---|---|---|---|---|
| $E_{u,a}(5.6)$ | -- | -12 | 0 | -5 | 0 | -26 |
| $E_{u,b}(5.6)$ | -12 | -- | 5 | 0 | -26 | 0 |
| $E_{u,a}(8.6)$ | 0 | -5 | -- | 32 | 0 | 182 |
| $E_{u,b}(8.6)$ | 5 | 0 | 32 | -- | -182 | 0 |
| $E_{u,a}(13.7)$ | 0 | -26 | -182 | 0 | -- | -229 |
| $E_{u,b}(13.7)$ | -26 | 0 | 0 | 182 | -229 | -- |

The atomic displacements along the $B_{2g}$ mode, induced by optical excitation of the two $E_u$ modes, were obtained by numerically solving the coupled system of nonlinear phononics equations.

Following the experimental conditions, we excite the CoF$_2$ crystal by an electric field pulse of the form $E(t) = E_0 e^{-\frac{t^2}{2w^2}}\cos(2\pi f_{IR} t)$, where $w$ = 0.3 ps, $f_{IR}$ = 13.7 THz, and $E_0$ is the field strength. Figure S2B and C show the resulting dynamics of the resonantly driven $E_u$ and nonlinearly coupled $B_{2g}$ modes for a field strength of 10 MV/cm. Clearly, driven oscillations of the $E_u$ mode rectify the lattice along the $B_{2g}$ mode coordinates. Figure S2D plots the of $B_{2g}$ mode distortion as a function of the excitation electric field strength, revealing super-linear growth as expected for the anharmonic excitation. Importantly, the induced amplitudes of the $B_{2g}$ mode (Fig. S2C) are orders of magnitudes larger than those achieved by applying static strain (Fig. S2A). We find $B_{2g}$ mode amplitudes as large as $Q_R$ = 40 $\sqrt{u}$ pm corresponding to a 4 pm displacement of fluorine ions. The largest experimentally applied pressure we found in the literature is 50 MPa [14], which according to our calculations induces an amplitude of only $Q_R$ = 0.06 $\sqrt{u}$ pm (fluorine ion displacements of 0.006 pm).

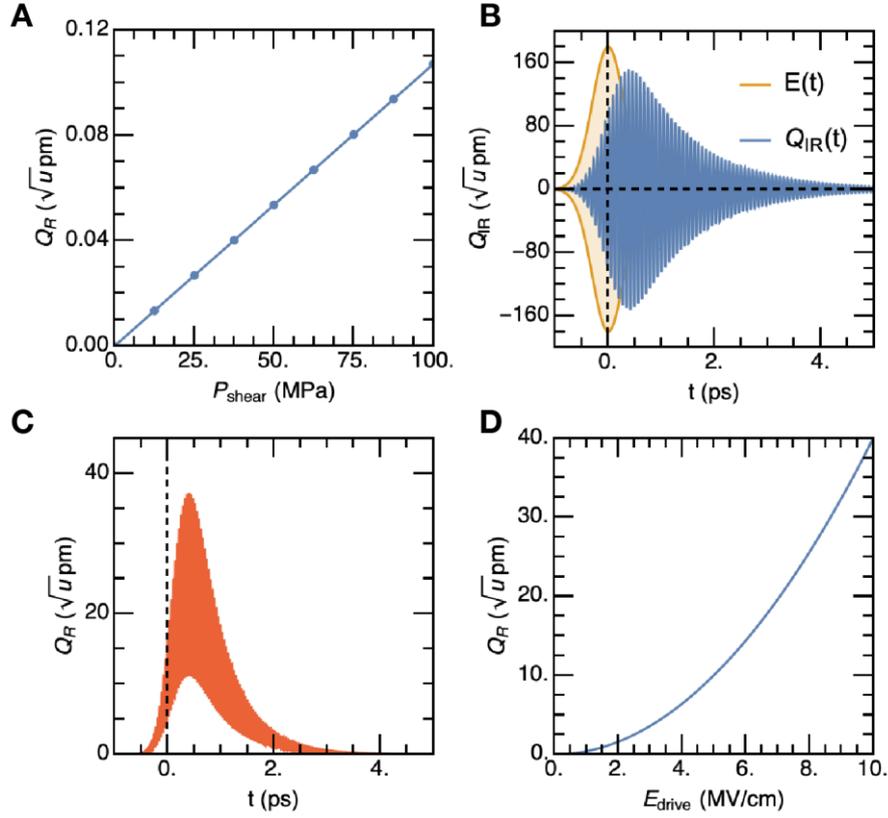

**Figure S3.** Structural distortion of the $Q_R = B_{2g}$ mode induced statically by pressure and dynamically by optical phonon excitation in $CoF_2$. In (A) we show the displacement of the $Q_R$ mode upon applying shear pressure. (B,C) displays the phonon dynamics of the $E_u$ and $B_{2g}$ modes after coherent excitation with an electric field pulse shown in orange. The pulse parameters are chosen to agree with the experimental settings. (D) shows the electric field dependence of the $Q_R = B_{2g}$ phonon mode amplitude.

### S3.4. Influence of distortions on magnetization

In the following we calculate the impact of the $B_{2g}$ mode structural distortion on the magnetic properties. We focus on two aspects: the change in orbital moment and the change in magnetic exchange interactions. For the former we utilize frozen phonon computations as described in Ref. [15] to explore changes in the spin $\langle S \rangle$ and orbital moment $\langle L \rangle$. We used the ELK code, which computes these quantities within a muffin tin sphere around an atom, which in our case of Co exhibits a radius of 1 Å.

The following Heisenberg Hamiltonian is used to describe the magnetic system in $CoF_2$

$$H_{\text{mag}} = J_a \sum_{\langle i,j \rangle} \hat{e}_i \cdot \hat{e}_j + J_b \sum_{\langle\langle i,j \rangle\rangle} \hat{e}_i \cdot \hat{e}_j + D_1 \sum_{[i]} (e_{i,z})^2 - D_2 \sum_{[j]} (e_{j,z})^2, \tag{S2}$$

where $J_1$ and $J_2$ are the nearest and next-nearest neighbor magnetic exchange coefficients, and $\langle i,j \rangle$ and $\langle\langle i,j \rangle\rangle$ denote the respective sums over nearest and next-nearest neighbor sites. $K_i$ represents the single-ion anisotropy of each Co site and the sums $[i]$ and $[j]$ run over the two different Co sublattices. The unit vectors $e_i$ represent the directions of the local moments on each site. This Hamiltonian is the same as that used in Ref. [16], where we replace the spin-half operators by unit vectors.

For our calculated equilibrium antiferromagnetic structure, the two Co sites have total magnetic moments $m_1 = \langle m_S \rangle + \langle m_L \rangle$ and $m_2 = -m_1$. We obtain values of

$$\langle m_S \rangle = 2.616 \ \mu_B$$

$$\langle m_L \rangle = 0.581 \ \mu_B \ .$$

Distorting the system along the eigenvector of the $Q_R = B_{2g}$ mode modulates the crystal field, resulting in changes of the orbital moment $\Delta\langle m_L \rangle$ on each site, as depicted in Fig. 1B in the main text. In contrast, the spin moments $\langle m_S \rangle$ remain unchanged by the distortion. The stronger sensitivity of the orbital magnetic moment to the crystal field is discussed in Ref. [15].

The total magnetic moments on the two sites become $m_1 = \langle m_S \rangle + \langle m_L \rangle + \Delta\langle m_L \rangle$ and $m_2 = -(\langle m_S \rangle + \langle m_L \rangle) + \Delta\langle m_L \rangle$. A finite net magnetic moment per unit cell thus appears with magnitude, $M = m_1 + m_2 = 2\Delta\langle m_L \rangle$, as shown in Fig. S4A. For our considered range of $Q_R < 40 \ \sqrt{u}$ pm, we obtained a linear dependence of $M = \left(0.405 \times 10^{-2} \frac{\mu_B}{\sqrt{u}\text{Å}}\right) Q_R$. Combining this change of magnetic moment with the pressure dependence of the $Q_R$ amplitude, we estimate the induced net magnetization by shear stress to $4.33 \times 10^{-3}$ $\mu_B$/GPa, close to the experimental value $6.2 \times 10^{-3}$ $\mu_B$/GPa [14]. Using this coefficient also for the phonon driven ferrimagnetic polarization, we obtain a magnetic moment of 0.17 $\mu_B$ per unit cell for the highest pump electric field strength of 10 MV/cm, which compares well with the experimentally determined value of 0.21 $\mu_B$ per unit cell (see Sec. S2).

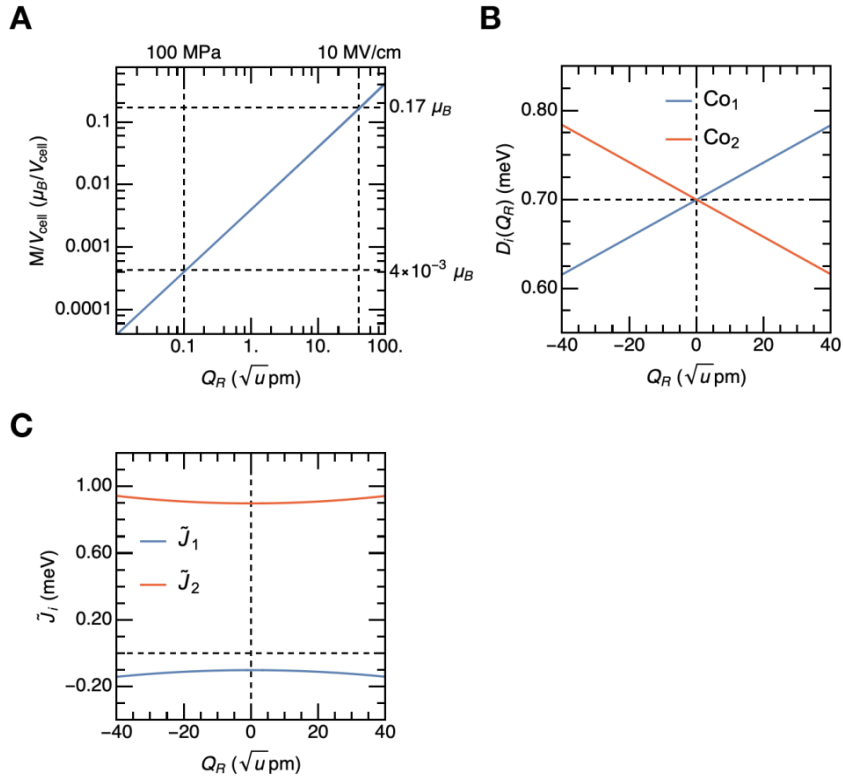

**Figure S4.** Dependence of the magnetic properties of $CoF_2$ upon distorting the system with the $Q_R = B_{2g}$ mode. In (A) we show the change total orbital magnetization per unit cell on a double logarithmic scale. The gridlines indicate exemplary values for applied shear stress and field strength of THz pulses.

Finally, we discuss changes of the magnetic interactions due to the $Q_R$ mode distortion. To map out the exchange and anisotropy coefficients $J_i$ and $D_i$ we use spin-flip calculations for each exchange interaction (see Ref. [17]). For our ground-state $CoF_2$ structure, we find $J_1 = -0.1$ meV, $J_2 = 0.9$ meV and $D_i = D_j = 0.7$ meV, in agreement with experimental findings [16, 18].

We then applied frozen phonon computations to determine the induced modifications of the exchange and single-ion anisotropy upon distorting the $Q_R = B_{2g}$ mode. We repeated the previous set of computations, taking particular care of the two different sublattices. As shown in Fig. 2B and 2C, the single-site anisotropy scales linearly with the amplitude of the $B_{2g}$ distortion, changing sign with the sign of the phonon amplitude. In contrast, the nearest and next-nearest neighbor exchange interactions scale quadratically as a function of the $B_{2g}$ phonon amplitude.

### S3.5. Computational settings

For our computations, we utilized the implementation of DFT employing the linearized augmented-plane wave method (LAPW) within the ELK-code [17]. As an approximation for the exchange-correlation functional we use the local spin density approximation (LSDA), which we augment with the Hubbard *U-J* parameter to account for the localized nature of the *d*-states of Co. According to Ref. [18] the magnetic properties, specifically the crystalline magnetic anisotropy, of $CoF_2$ are very sensitive to the values of *U* and *J*. We use *U* = 4 eV and *J* = 0.6 eV. For this combination of values, we reproduce the measured equilibrium values for magnetic exchange and uniaxial anisotropy. As numerical settings for our computation we use ELK in the very high-quality mode, which sets the maximum cutoff of the plane waves times the average muffin-tin radius $R_{MT}G_{max}$ = 9.0. We increased the *k*-mesh to 14×14×20 points to sample the Brillouin zone of the unit cell; all other parameters are kept in the very high-quality setting. Note that for computations of larger cells we reduced the *k*-mesh according to the increase of the cell size. Last, for all computations regarding the orbital magnetism and magnetic anisotropy, we include the spin-orbit interaction, as implemented in ELK, in our computations.

## S4. Possible model for magnetization dynamics

In the following, we develop a phenomenological model to describe the coupled dynamics of the $B_{2g}$ phonon distortion, represented by its amplitude $Q_R$, the antiferromagnetic order, and the ferrimagnetic moments.

We define the antiferromagnetic order parameter as the staggered magnetization

$$L = (m_1 - m_2)$$

and the induced ferrimagnetic order parameter as the magnetic polarization

$$M = (m_1 + m_2),$$

where $m_i$ are the *z*-components of the magnetic moments on the two Co sublattices. Considering the $D_{4h}$ (4/mmm) point group of $CoF_2$, we find symmetries $L \in A_{2g}$, $M \in B_{1g}$, and $Q_R \in B_{2g}$.

### S4.1 Phonon-piezomagnetic free energy

Following Ref. [19], the free energy density of the system, including all symmetry allowed terms up to the fourth order, reads as

$$\begin{aligned}\mathrm{F}(L, M, Q_R) =\ &\frac{a_L}{2}L^2 + \frac{b_L}{4}L^4 + \frac{a_M}{2}M^2 + \frac{b_M}{4}M^4 + \frac{a_R}{2}Q_R^2 + \frac{b_R}{4}Q_R^4 + \lambda Q_R L M \\ &+ c_1 L^2 M^2 + c_2 L^2 Q_R^2 + c_3 Q_R^2 M^2.\end{aligned} \quad (S3)$$

The coefficients $a_L$, $a_M$ and $a_R$ parameterize the harmonic energy potential for *L*, *M*, and $Q_R$, respectively, $b_L$, $b_M$ and $b_R$ describe their anharmonicities, $\lambda$ is the trilinear coupling coefficient, and $c_1$, $c_2$ and $c_3$ are higher order coupling coefficients.

In the $CoF_2$ ground state, *L* is the primary order parameter with the finite equilibrium value,

$$L_0 = \sqrt{-\frac{a_L}{b_L}} = m_1^0 - m_2^0. \qquad (S4)$$

To ensure stability of $L_0$, it is required that $a_L < 0$ and $b_L > 0$. In addition, since the equilibrium values $M_0 = 0$ and $Q_R = 0$ in the ground state, the Hessian matrix of $F(L,M,Q_R)$ must be positive definite with respect to the magnetic polarization $M$ and phonon amplitude $Q_R$, leading to the constraints $a_M > 0$ and $a_R > 0$.

As discussed in Sec. S3.4, the phonon-piezomagnetic coupling considered here modifies the magnetic moments on the two sublattices such that $m_1 = m_1^0 + \delta m/2$ and $m_2 = m_2^0 + \delta m/2$ for small $Q_R$ amplitudes. The magnetic order parameters become modified as follows

$$L = \left(m_1^0 + \frac{\delta m}{2}\right) - \left(m_2^0 + \frac{\delta m}{2}\right) = L_0$$

and

$$M = \left(m_1^0 + \frac{\delta m}{2}\right) + \left(m_2^0 + \frac{\delta m}{2}\right) = \delta m,$$

where we have taken into account that $m_1^0 + m_2^0 = M_0 = 0$. Thus, since $L$ remains constant by this physical mechanism, the last two coupling terms in Eq. S3, which involve $L^2$ coupling to the ferrimagnetic moment and to the phonon displacement, can be regarded as rigid energy shifts and can be absorbed into the coefficients $|a_M|$ and $|a_R|$. Furthermore, since the auxiliary order parameters $Q_R$ and $M$ appear as small perturbations of the ground state, we eliminate terms higher than quadratic order in these parameters.

In this case, the free energy density becomes

$$F(L, \delta m, Q_R) \approx -\frac{|a_L|}{2}L^2 + \frac{|b_L|}{4}L^4 + \frac{|\widetilde{a_M}|}{2}\delta m^2 + \frac{|\widetilde{a_R}|}{4}Q_R^2 + \lambda Q_R L \delta m, \qquad (S5)$$

where $\widetilde{a_M} = a_M + c_1 L_0^2$ and $\widetilde{a_R} = a_R + c_1 L_0^2$.

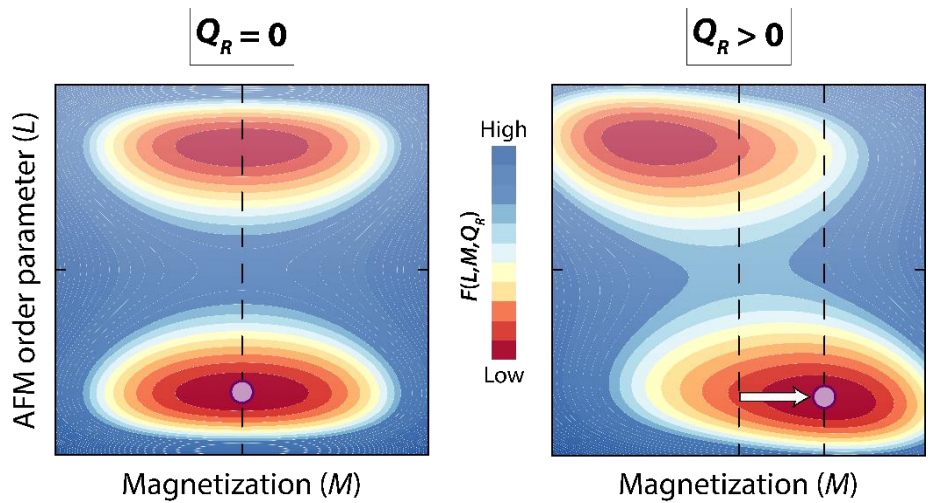

**Figure S5.** Free energy of $CoF_2$ as a function of the antiferromagnetic and ferrimagnetic order parameter and the distortion $Q_R$ from Eq. S5. The left panel illustrates the equilibrium case with a double well potential for $L$ and single well for $M$. The purple dot shows one of the absolute minima of the potential landscape corresponding to the ground state in one antiferromagnetic domain. (right) Displacing the $B_{2g}$ symmetry phonon $Q_R$ induces a shift of the minima due to the trilinear coupling term $\lambda Q_R L M$ and a new ground state with finite $M$ becomes preferred.

This free energy landscape is illustrated in Fig. S5. At equilibrium, the free energy has two minima at opposite values of the antiferromagnetic order (+$L_0$ and -$L_0$), but at zero net magnetization, corresponding to the two possible antiferromagnetic domains. For a finite distortion of $Q_R$, the two minima positions shift to finite values of $M$, resulting in a ferrimagnetic polarization, but with different signs of $M$ for the two different domains.

Finally, for considering the non-equilibrium behavior of $M$, we assume that the dynamics of $Q_R$ are fast relative to $M$ (i.e. $|a_R| \gg |a_M|$). So, for a given time-dependent phonon distortion $Q_R(t)$, the energy density in terms of the induced magnetization reads as

$$F_M \approx \frac{|\widetilde{a_M}|}{2}\delta m^2 + \lambda Q_R(t) L \delta m, \tag{S6}$$

From this expression, it can be seen that the magnetization is governed by a harmonic energy potential and is driven linearly by $Q_R$.

### S4.2 Dynamics of induced magnetization

We consider here the free energy described above as an effective potential to understand the non-equilibrium dynamics of the magnetization, as previously applied to magnetoelastic media [20, 21]. This simplified potential provides only a longitudinal effective field; hence, the directions of the moments within the two antiferromagnetic sublattices remain constant and only their relative magnitudes vary. In analogy to the known inertial dynamics of the order parameter in two-sublattice antiferromagnets [22, 23], we write the equation of motion for $M$ as

$$\ddot{M} + 2\zeta\dot{M} = -\gamma\Omega M_a \frac{\partial F_M}{\partial M}, \tag{S7}$$

which becomes upon plugging in $F_M$ from Eq. (S5),

$$\ddot{M} + 2\zeta\dot{M} + \omega^2 M = -q\lambda L Q_R(t), \tag{S8}$$

with the phenomenological damping coefficient $\zeta$ and scaling factor $q$. The term $\omega = \sqrt{q|\widetilde{a_M}|}$ represents the effective frequency of this magnetization mode.

To simulate the time dependence of $M(t)$ in our experiment, we incorporate the driving term $Q_R(t)$ from the nonlinear phonon dynamics described in Sec. S3.3 and the equation of motion for $L(t)$, equivalent to Eq. (S8), subject to the initial condition for $L_0$ from Eq. (S4). The resulting $M(t)$ evolution is shown in Fig. 6B in the main text for different incident electric field strengths. The parameters in the equation of motion for $M(t)$ were adjusted to best match the experimental data, yielding $\frac{1}{\zeta} \approx 310$ ps, $\frac{\omega}{2\pi} \approx 0.46$ GHz, and $q\lambda \approx 220 \frac{\text{GHz}^2}{\sqrt{u}\text{Å}}$. Figure S6 shows the time dependence of $M$, $L$, $Q_R$, and $Q_{IR}$ upon driving by an electric field pulse with 12 MV/cm peak strength, 500 fs FWHM, and 12 THz center frequency.

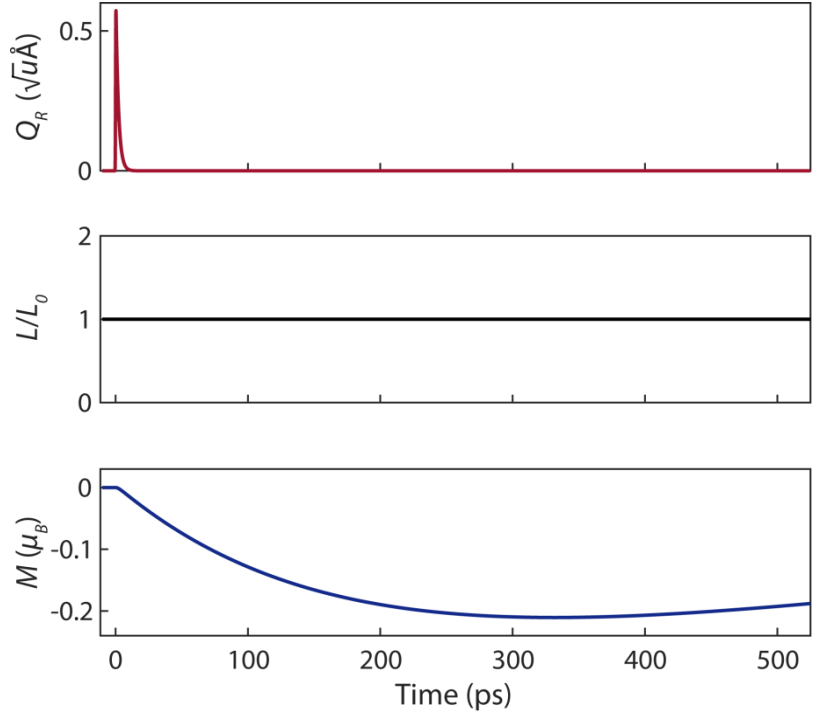

**Figure S6.** Coupled dynamics of the $B_{2g}$ phonon displacement, antiferromagnetic order parameter, and magnetization upon resonant infrared phonon driving. The motion of $Q_R$ was determined by the nonlinear phononic coupling (see Sec. S3.3), while the dynamics of $L$ and $M$ were numerically calculated from Eq. S8. The parameters used for this calculation are given in the text.

The behavior of the magnetization beyond ~10 ps is that of an overdamped harmonic oscillator driven by a short impulse. It is useful to consider the form of the solution to identify the relevant physical processes and associated time constants, which in the overdamped regime can be written as

$$M(t) = \kappa Q_0 \tau_D \left(1 - e^{-t/\tau_D}\right) e^{-t/\tau_R}. \tag{S9}$$

Here, $Q_0$ expresses the dependence of the induced magnetization on the amplitude of the $Q_R$ displacement, and $\tau_D$ and and $\tau_R$ represent the damping and relaxation time constants, respectively. The constant $\kappa$ is related to the inertial component of the magnetic system, which gives rise to the growth of $M$. This initial impulse is damped over the time $\tau_D$, after which the magnetisation would remain constant in the absence of relaxation. Relaxation to the ground state takes place over a longer time $\tau_R$, which we may identify with the spin-lattice relaxation time. Based on the numerical solution from Fig. S6, we obtain $\tau_D \approx 150$ ps and $\tau_R \approx 1200$ ps. This longitudinal relaxation rate agrees with estimates from theory and measurements of magnetic critical scattering [24, 25], as well as the typical spin-lattice equilibration rates in magnetic insulators [26, 27].

### S4.3 Discussion

The proposed model is purely phenomenological, and the observed magnetization dynamics may arise from a number of possible non-equilibrium scenarios. The damped oscillation seen in the experiment does not correspond to an equilibrium spin wave: the lowest magnon frequency in $CoF_2$ is ~1 THz [26]. Rather, as noted above, a likely explanation for the time dependence may rather arise from the interplay of inertial antiferromagnetic dynamics and longitudinal spin relaxation. The dynamics of the magnetic moments in a two-sublattice antiferromagnet such as $CoF_2$ are most appropriately described by coupled Landau-Lifshitz-Bloch (LLB) equations, as formulated in Refs. 28, 29. The phonon-piezomagnetic driving term in the free energy (Eq. S6) would provide an effective

(longitudinal) field in this context, in addition to the equilibrium anisotropy and exchange fields. This effect might also be represented by a prompt unbalancing of the *g*-factor between the two sublattices. The perturbation hence excites the magnetic system out of equilibrium, which then evolves according to the inertial and relaxation time scales described above.

An additional contribution to the dynamics is the time dependence of the anisotropy and exchange on each site due to the phonon displacement (as shown in Fig. S4). In the distorted state, the degeneracy between the magnon bands is lifted, raising the possibility that the observed magnetization oscillation is the result of a beating between the two modes. Another possibility is that the oscillation corresponds to an amplitude (or "Higgs") mode of the magnetic polarization stimulated by the symmetry-breaking phonon, which would be consistent with the free energy description above.